# Microwave-assisted synthesis of LaMnO$_{3+d}$: Tuning physical properties with microwave power

Marimuthu Manikandan, Arup Ghosh, and Ramanathan Mahendiran[*]

*Department of Physics, National University of Singapore, 2 Science Drive 3, Singapore 117551, Republic of Singapore*

**Abstract**

Synthesis of transition metal oxides by microwave irradiation is a faster and energy-saving method compared to conventional heating in an electrical furnace because microwave energy is directly converted into heat within precursors. However, not much is known about how the physical properties are modified by the power of microwaves. We synthesized LaMnO$_{3+d}$ by irradiating oxide precursors with microwaves and studied the impact of microwave power ($P$ = 1000 W, 1200 W, 1400 W and 1600 W) on magnetism, resistivity, magnetoresistance, thermopower, magnetic entropy change, magnetostriction, and electron spin resonance. It is found that paramagnetic to ferromagnetic transition becomes sharper, saturation magnetization increases and electrical resistivity at low temperatures dramatically decreases as $P$ increases. While the resistivity of samples irradiated with MW power of $P \leq 1400$ W show insulating-like behavior down to 50 K, an insulator-metal transition occurs in the sample exposed to $P$ = 1600 W and this sample also shows a maximum magnetoresistance (= -55%), magneto-thermopower (=-87%), magnetostriction (-180 x10$^{-6}$) for $H$ = 50 kOe and magnetic entropy change of 4.78 J/kg. K for $\Delta H$ = 30 kOe around the Curie temperature. The intensity of electron spin resonance spectra at 300 K increases with $P$. We postulate that the much enhanced physical properties observed for $P$ = 1600 W sample arise from the creation of higher hole density, chemical homogeneity and increased grain size. Our study shows that microwave power can be used as a knob to tune magnetism and other physical properties to our advantage.

[*]Author for correspondence: R. Mahendiran (email: *phyrm@nus.edu.sg*)

# 1. Introduction

LaMnO$_3$ is the parent compound of the colossal magnetoresistive oxide family known as manganites [1]. It is an antiferromagnetic insulator at low temperatures with a Neel temperature $T_N$ = 140 K but turns into a ferromagnetic metal upon partial substitution of divalent alkaline earth ion (A) for La$^{3+}$ cation as in La$_{1-x}$A$_x$MnO$_3$ (A= Sr$^{2+}$, Ca$^{2+}$ and Ba$^{2+}$) which oxidizes a fraction of Mn$^{3+}$ ($d^4$) into Mn$^{4+}$ ($d^3$) to maintain charge neutrality. Thus, $e_g$-holes (Mn$^{4+}$:$t_{2g}^3 e_g^0$) are doped in the parent compound [2]. A strong interatomic Hund's coupling ties the direction of the spin of $e_g$ hole/electron to $t_{2g}^3$ core spin at the atomic level. Zener's double exchange interaction mediates ferromagnetic interaction between Mn$^{3+}$ and Mn$^{4+}$ cations via mobile $e_g$ holes [3]. The ferromagnetic transition temperature increases with the Mn$^{4+}$/Mn$^{3+}$ ratio and reaches the maximum ($T_C$ = 372-376 K) for $x$ = 0.33 in La$_{1-x}$Sr$_x$MnO$_3$ and the $T_C$ for a fixed hole content ($x$) decreases with decreasing size of the divalent cation or rare earth cation. Ferromagnetism in LaMnO$_3$ can also be induced by the substitution of monovalent cations (eg. Na$^{1+}$ and K$^{1+}$) for La$^{3+}$ which creates twice the hole density compared to the divalent cations substitution [4,5,6] and the $T_C$ reaches a maximum value of 326 K for $x$ = 0.2 in La$_{1-x}$Na$_x$MnO$_3$ series [7,8].

Interestingly, ferromagnetism and metallic-type resistivity were also reported in LaMnO$_{3+d}$ with excess oxygen [9,10,11,12,13]. Since excess oxygen does not go to interstitial site in the perovskite structure, it can be considered as cation-deficient La$_{1-d}$Mn$_{1-d'}$O$_3$ [14]. Arulraj et al. [15] studied the influence of a non-equal proportion of La and Mn-site cation vacancies on magnetic, electrical properties and found that Mn-site deficiency was detrimental to metallic and ferromagnetic properties compared to the La-site deficiency [15]. LaMnO$_{3+d}$ has been traditionally synthesised using either the solid state reaction route at $T$= 1000 °C-1100 °C) and high oxygen pressure [16] or by a wet chemical route at lower temperatures ($T$ = 800 - 900 °C) or electrochemical oxidation [9,15]. Regaieg et al. [17] reported rapid synthesis of LaMnO$_3$ by spark plasma sintering and found that T$_N$ barely changed with respect solid state synthesized sample. However, post annealing in an electrical furnace at 1000 °C for 10 hrs led to induction of Ferromagnetic transition at $T_C$ ~ 241 K. Wang and Saito reported synthesis of LaMnO$_3$ by high energy ball milling La$_2$O$_3$ and Mn$_2$O$_3$ at room temperature but did not study its physical properties [18]. Irradiation of polycrystalline LaMnO$_3$ by fast neutron with energy more than 1 MeV reduced Jahn-Teller distortion and transformed the low temperature spin configuration antiferromagnetic to canted antiferromagnetic [19]. While the zero-field resistivity of the oxygen excess $LaMnO_{3.15}$ showed semiconducting-like behavior 10 K with a weak anomaly around 210 K, high energy electron beam irradiation induced insulator to metal transition at T= $T_{IM}$ which shifted downwards ($T_{IM}$ = 175 to ~130 K) with increasing dosage level of electron beam [20]. In addition, the magnitude of the resistivity at its peak value also increased with the dosage level. In this context, it will be interesting to investigate the effect microwave irradiation on physical properties.

Gibbson et al. [21] successfully synthesized $La_{0.97}MnO_3$, $La_{0.7}Sr_{0.3}MnO_3$ and $La_{0.5}Ca_{0.5}MnO_3$ ceramics using 950 W microwave oven and studied their structural, electrical resistivity and magnetoresistance properties. In microwave-assisted synthesis, microwave radiation interacts with reactants and microwave energy is dissipated as heat inside reactants when electrical dipoles (polar molecules) in the reactants are unable to oscillate in phase with the electric field component of the MW radiation. Microwave heating is instantaneous, it heats the reactants to 1000 °C and above within a few minutes to tens of minutes. However, at least one of the reactants should be a good microwave absorber for heating to be initiated. Rare earth and alkaline earth metal salts with hydroxide ions ($OH^-$) and nitrates ($NO_3^-$) were used earlier [22,23]. In the present work, we used microwave radiation to synthesize $LaMnO_3$ directly from their respective oxide precursors with MW power ($P$) as a knob to tune its physical properties. Calcination and synthesis of the sample were done in a microwave furnace. We fixed the final sintering temperature to 1000 °C and varied MW power ($P$ = 1000 W, 1200 W, 1400 W and 1600 W) to reach that temperature. The MW synthesized samples with different $P$ were studied for physical properties such as structure, magnetic, electrical and thermal transport, magnetostriction, magnetic caloric effect and electron paramagnetic resonance.

## 2. Experimental Details

$LaMnO_{3+d}$ samples were prepared from the stoichiometric ratio of dehydrated $La_2O_3$ and $Mn_2O_3$ mixture by heating them in a multimode muffle microwave furnace (Milestone Inc, Italy, PYRO model MA 194-003) operating at 2.45 GHz. The precursors were mixed for 12 g using mortar and pestle and divided into six equal portions. Initially, three portions were separately irradiated by MW with a power $P$ = 1000 W for 10 min at the set temperatures of 800 ºC, 900º C and 1000 ºC, respectively. Then, the irradiated powder was ground and made into individual circular pellets and sintered them separately to the above-set temperatures and soaked there for 20 min for phase formation. X-ray diffraction pattern indicated unreacted $Mn_2O_3$ in samples prepared for T < 1000 °C. Hence, further syntheses were carried out only at 1000 °C with different MW powers ($P$ = 1200 W, 1400 W and 1600 W) using the rest of the portions of the mixture and followed the same procedures mentioned above. The temperature was set to reach 1000 °C in 10 min (heating rate of ~ 100 °C/min) and the pellet was soaked (sintered) at that temperature for 20 min. Then, the MW power was switched off and the pellet was cooled down to room temperature in 100 min (cooling rate of ~ 12 °C/min). A portion of each pellet was crushed into powders to do X-ray diffraction for structural analysis. Field emission electron microscopy (FE-SEM) was used to image the morphological features. Magnetization was measured using a vibrating sample magnetometer probe attached to a physical property measurements system (PPMS). Isothermal field-sweep up to 50 kOe was measured at 5 K intervals over a certain temperature range around the Curie temperature to estimate magnetic entropy change. Temperature-dependent electrical resistivity ($\rho$) and longitudinal thermopower ($S$) were simultaneously measured using a custom-designed sample stage built on a standard puck for the PPMS. A commercial broadband ferromagnetic resonance spectrometer (NanOsc Phase-FMR from Quantum Design Inc.) that uses a coplanar waveguide to magnetically excite the sample placed on it was used for magnetic resonance studies. Superconducting coils in the PPMS provide a *dc* magnetic field that is transverse to the microwave (MW) magnetic field created in the coplanar waveguide by the flow of MW current. Further, the *dc* field is modulated by a low-frequency *ac* magnetic field using a pair of Helmholtz coil to increase the signal-to-noise ratio. The resonance spectra are recorded as the field derivation of MW power absorption (*dP/dH*). Magnetostriction along the direction of the applied *dc* magnetic field and in the temperature range from 300 K to 10 K was measured using a capacitance dilatometer probe inserted in the PPMS. A polished cube-shaped sample of size 2 x 2 x 2 mm was placed between two circular capacitive electrodes in the dilatometer probe. The change in the length of the sample was measured through a change in the capacitance of the dilatometer using a high-resolution capacitance bridge (Andeen Hagerling, model AH2500A).

## 3. Results and Discussion

### 3.1. Structural and Morphological analysis

Figure 1(a) shows X-ray diffraction patterns of $LaMnO_3$ samples synthesized at 1000 °C using different microwave powers ($P$ = 1000 W, 1200 W, 1400 W and 1600 W) from the bottom to the top panel. We also synthesized LaMnO$_3$ at 800 °C and 900 °C using $P$ = 1000 W but they showed more unreacted Mn$_2$O$_3$. The intensity of the unreacted/secondary peaks were reduced as the temperature increases to 1000 °C and a single phase forms at 1000 °C when $P \geq 1200$ W. Rietveld refinement for all the samples could be fitted for rhombohedral structure ($R\overline{3}C$ space group) and the fitting are shown in the top panel of Fig. 1(a) for $P$ = 1600 W. There is not much variations in the lattice parameters for each sample and they are around $a = b = 5.5327(7)$ Å, $c = 13.3567(9)$ Å, $\alpha = \beta = 90°$, and $\gamma = 120°$. Depending on the concentration of Mn$^{4+}$, LaMnO$_{3+d}$ crystallizes in different structures: orthorhombic (Mn$^{4+} \leq 20\%$), rhombohedral (20% < Mn$^{4+} \leq 30\%$) and cubic (Mn$^{4+}$ > 30%) [9-15]. Hence, it is likely that our MW processed samples possess Mn$^{4+}$ between 20% and 30% which will be detailed in the magnetization and electrical transport properties.

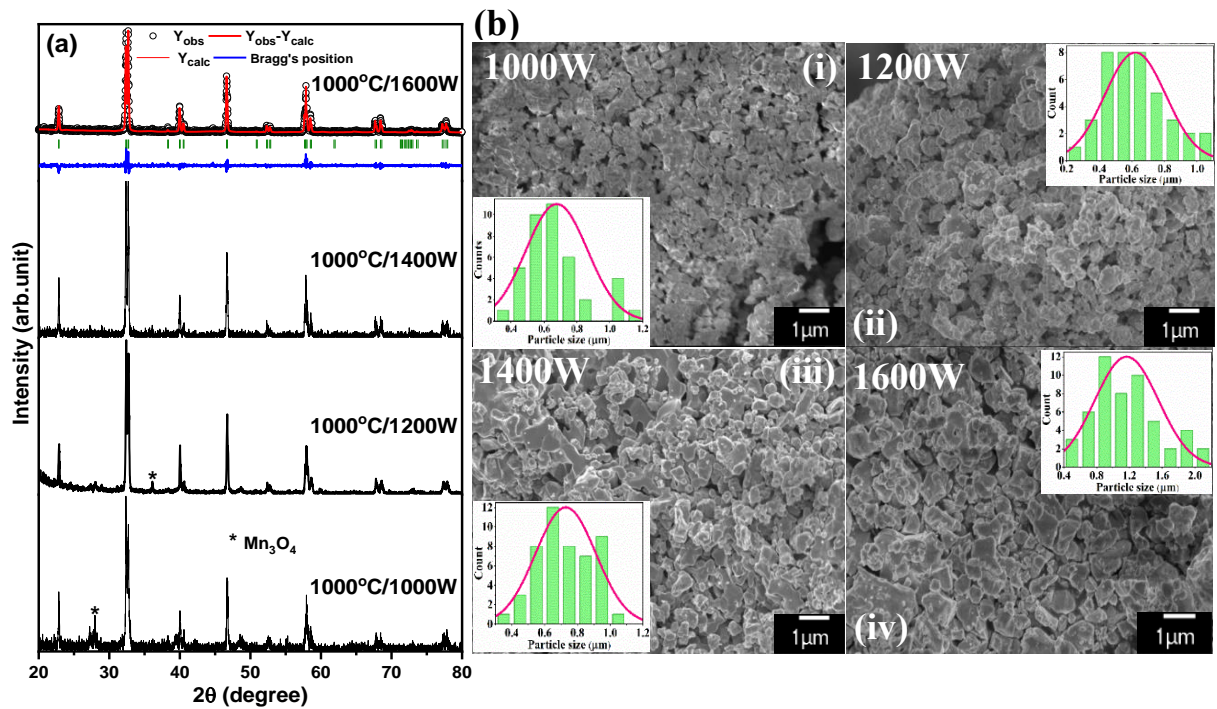

**Fig. 1.** X-ray diffraction of LaMnO$_3$ synthesised by microwave irradiation at 1000 °C with different microwave powers ($P$ = 1000 W, 1200 W, 1400 W and 1600 W). Rietveld fit for the sample prepared for $P$ = 1600 W is shown. The open black circles, red lines, blue lines and green vertical bars represent the experimental data, calculated pattern, difference curve and Bragg position, respectively.

Scanning electron microscopy (SEM) images (i)-(v) for LaMnO$_3$ prepared at $P$ = 1000 W to 1600 W are shown in Fig. 1(b). All the samples show agglomeration of spherical and irregular-shaped particles, and the agglomeration is reduced as the MW power is increased. Also, the spherical particles fuse with each other and become like stone pebbles with irregular shapes and sizes, which is more apparent for $P = 1600$ W. The particle size distribution in each sample is shown in the respective image. While all the samples consist of micron-ranged particles, the size of the particles increases with increasing power. At 1000 W, more particles are distributed in the range between 0.4 µm and 0.8 µm, and the particle population extends to 1 µm at 1200 W and 1400 W. The sample prepared at 1600 W shows relatively bigger particles and they are distributed in the wide size range between 0.4 µm and 2.2 µm.

## 3.2. Magnetization

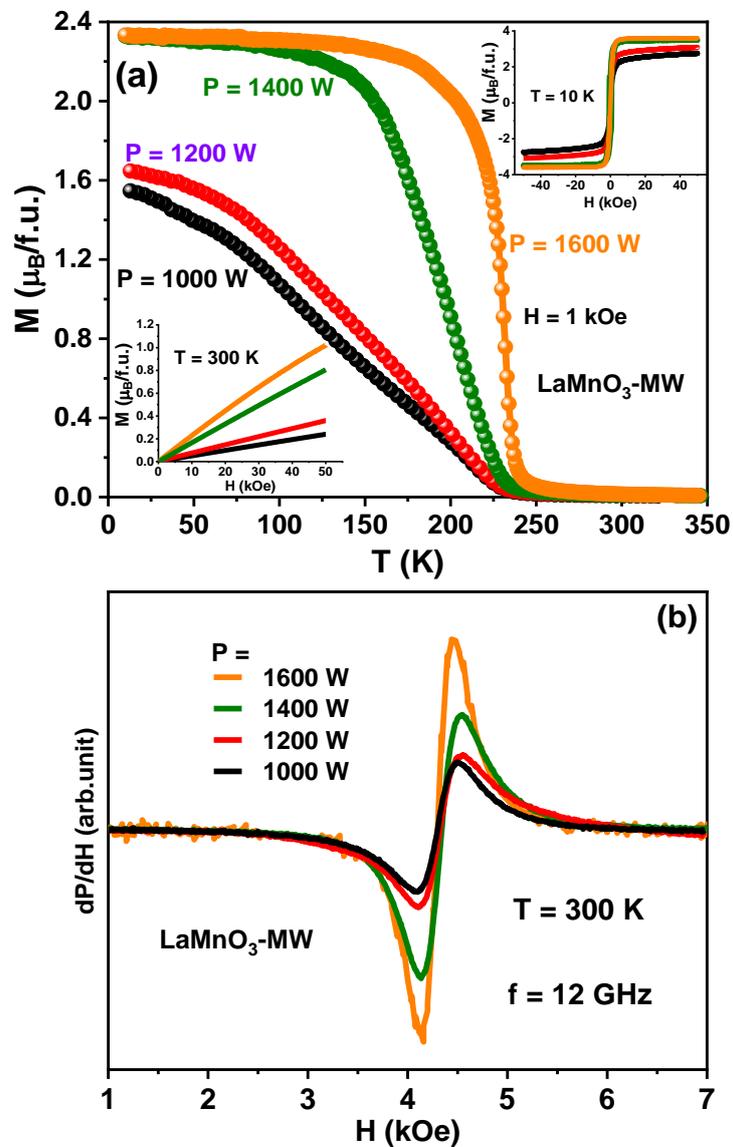

**Fig. 2. (a)** Temperature dependence of magnetization $M(T)$. Insets show field dependence of magnetization at 10 K (right top inset) and 300 K (left bottom inset) for LaMnO$_3$ synthesised at different MW powers. **(b)** Magnetic field dependence of the microwave power absorption ($\Delta P$) of LaMnO$_3$-MW at 300 K.

Fig. 2(a) shows the temperature dependence of magnetization $M(T)$, for LaMnO$_3$ samples prepared with different MW powers measured under $H = 1$ kOe. All the samples undergo paramagnetic to ferromagnetic transition as the temperature is lowered as indicated by rapid increase in $M(T)$. The paramagnetic-ferromagnetic transition is broad in the samples exposed to MW powers of $P = 1000$ W and 1200 W but becomes sharper for $P = 1400$ W and 1600 W samples. The Curie temperature $T_C$ deduced from the inflection point of $dM/dT$ increases from 190 K for $P = 1000$ W to 236 K for $P = 1600$ W. All the parameters obtained from magnetization are listed in Table I. The obtained $T_C$ are close to the value reported for LaMnO$_3$ with $R\bar{3}C$ space group (200 K -240 K) [9-15]. The right-top inset depicts $M(H)$ isotherms at 10 K up to $H = 50$ kOe for different MW powers $P$. The field dependence of $M$ is that of a soft ferromagnet and the value of $M$ at the highest field (50 kOe) increases with increasing MW power, from 2.76 $\mu_B$/f.u. for $P = 1000$ W to 3.62 $\mu_B$/f.u for $P = 1600$ W. Typically, a saturation magnetization of 3.6 $\mu_B$/f.u. is realized in alkaline earth-doped La$_{1-x}$Sr$_x$MnO$_3$ and La$_{1-x}$Ca$_x$MnO$_3$ for $x = 0.4$, i.e., with 40% of Mn$^{4+}$ doping [1]. $M(H)$ isotherms at 300 K shown on the left-bottom inset show paramagnetic behavior and the value of $M$ at 50 kOe increases with $P$. The increase of $M$ value in the paramagnetic state suggests that $e_g$-holes (Mn$^{4+}$ cation) are most likely doped with increasing MW power. To confirm this, we measured ESR spectra at $f = 12$ GHz and at room temperature for different MW power and they are shown in Fig. 2(b). Each sample exhibits a single Lorentizan spectrum and the amplitude of the peak increases and line width ($\Delta H$) of the peak decreases with increasing MW power. The intensity of the absorption are directly proportional to the concentration of active paramagnetic entities in the sample. The increasing intensity of the ESR signal with MW power suggest that more magnetic entities contributing to the ESR signal increases, i.e., Mn$^{4+}$ content increases. This is in agreement with the results of Tover et al, who found that doubly integrated intensity of ESR signal in LaMnO$_{3+\delta}$ ( $0 \leq \delta \leq 0.07$) at room temperature increased with excess oxygen [24] We have also fitted the inverse susceptibility above $T_C$ with the Curie-Weiss law $\chi^{-1} = C/(T-\theta)$ (not shown here) where $C$ is the Curie Constant which is related to the effective magnetic moment ($\mu_{eff}$) via $\mu_{eff} = 2.83\sqrt{C}$ and $\theta$ is the Weiss constant. The Weiss constant is positive for all the samples which indicates the ferromagnetic correlations in the paramagnetic state. While $\theta$ are close to each other for $P$ = 1000 and 1200 W samples, it increases for higher powers. $\mu_{eff}$ also increased with $P$ from 5.48 $\mu_B$ for $P$ =1000 W to 6.4 $\mu_B$ for $P$ = 1600 W sample. The enhanced $\mu_{eff}$ in the paramagnetic can come from an increase in the hole density.

The mixed powder of precursors are heated internally and instantaneously upon absorbing microwave radiation. The higher the MW power, higher is the heating rate and chemical reaction among precursors takes place in a non-equilibrium thermodynamic state. It is most likely possible that La and Mn vacancies are created under rapid heating and the cation vacancies increase with increasing microwave power. To sustain charge neutrality in La$_{1-x}$Mn$_{1-x}$O$_3$, $x$ fraction of Mn$^{4+}$:$t_{2g}^3 e_g^0$ are created. Zener's double exchange interaction between created Mn$^{4+}$ and Mn$^{3+}$ leads to ferromagnetism. Rhombohedral structure does not support static Jahn-Teller distortion.

| MW power (P) | $T_C$ dM/dT (K) | M at 50 kOe | $\theta$ (K) | $\mu_{eff}$ ($\mu_B$) |
|---|---|---|---|---|
| 1000 W | 190 | 2.76 | 202 | 5.48 |
| 1200 W | 195 | 2.82 | 205 | 5.89 |
| 1400 W | 224 | 3.46 | 225 | 6.04 |
| 1600 W | 236 | 3.62 | 236 | 6.4 |

**Table. 1.** Magnetization parameters: Curie transition temperature ($T_C$), magnetization at maximum field ($M_{max}$), Curie Weiss constant ($\theta$) and effective magnetization ($\mu_{eff}$) for the samples prepared with different MW powers.

### 3.3. Resistivity and Thermopower

Fig. 3(a)-(d) shows the temperature dependence of *dc* resistivity ($\rho$) for samples exposed to different MW Powers. $\rho(T)$ were measured under $H$ = 0 kOe and $H$ = 50 kOe for all the samples and also at two intermediate fields ($H$ = 10 and 30 kOe ) over a certain temperature range in some samples. It is remarkable that while $\rho(T)$ of the samples for $P$ = 1000 W, 1200 W and 1400 W show insulating-like behaviour (d$\rho$/d$T$ > 0) down to 50 K, the sample exposed to $P$ = 1600 W undergoes an insulator to metal transition at a temperature ($T_{IM}$) very close to $T_C$. The magnitude of $\rho$ at 50 K is dramatically decreased with increasing MW power, from 2.5 x 10$^4$ $\Omega$ cm for $P$ = 1000 W sample to 0.6 $\Omega$ cm for $P$ = 1600 W sample. The applied magnetic field influences resistivity differently in both samples. While the resistivity on the low-temperature

side (T < 150 K) decreases dramatically with increasing strength of magnetic field for *P* = 1000 W sample (also likely for *P* = 1200 & 1400 W samples), the decrease is prominent around $T_C$ for *P* = 1600 W sample. The resistivity peak for this sample decreases and shifts to high temperature as *H* increases as generally found in the well sintered samples prepared by conventional heating.

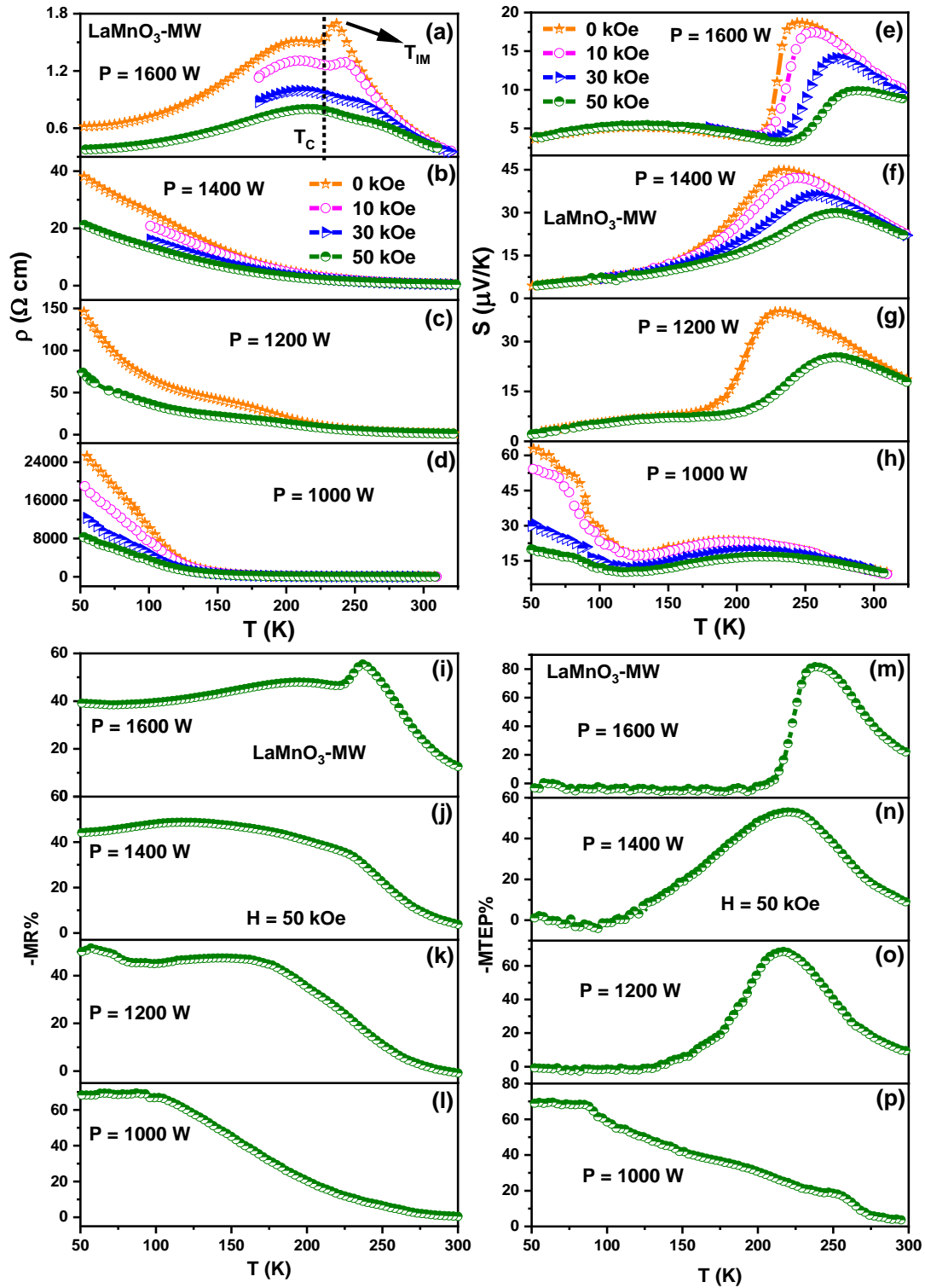

**Fig. 3.** Temperature dependence of resistivity ($\rho$) **(a-d)**, thermoelectric power (*S*) **(e-h)** under different magnetic fields for MW-LMO sample prepared with different MW powers and magnetoresistance (MR %) **(i-l)** and Magnetothermopower (MTEP %) **(m-p)** for *H* = 50 kOe.

Fig. 3(e)-(h) show the temperature dependence of the thermopower *S* in zero and non-zero magnetic fields for the samples corresponding to those in Fig. 3(a)-(d). The sign of *S* is positive in the entire temperature range measured (350 K – 50 K) which indicates that holes are the majority charge carriers in these compounds. The insulator-metal transition for *P* = 1600 W is also accompanied by a change in the nature of *S* (*T*, *H* = 0) from semiconducting-like (*S* increasing with lowering temperature) in the paramagnetic state to an abrupt decrease upon the onset of ferromagnetism, culminating in a peak at $T_C$. The peak decreases in magnitude and shifts upwards in temperature with increasing *H* similar to the resistivity peak. However, *S* for *T* << $T_C$ is least affected by the applied magnetic field in contrast to the resistivity. It is interesting to note that *S* (*T*, *H* = 0) also exhibits a peak around $T_C$ for *P* = 1400 and 1200 W samples even though $\rho(T, H =0 )$ does not show. The abrupt decrease of *S* just below $T_C$ is spread over a wider temperature range compared to *P* = 1600 W but the peak shifts towards high-temperature and decrease in magnitude under an external magnetic field are similar to the *P* = 1600 W sample. On the other hand, *S*(*T*, *H* = 0) for *P* = 1000 W shows a broad hump around 200 K and a tendency to increase rapidly below 100 K. Unfortunately, *S* is not measurable below 50 K due to the high two probe resistance of the sample. *S*(*T*) below 100 K is severely affected by the magnetic field in the 1000 W sample compared to high temperature, and this behavior is very different from samples irradiated with other MW powers.

Fig. 3(i)-(l) shows the temperature dependence of magnetoresistance (*MR*) for *H* = 50 kOe. Fig. 3(m)-(p) shows magneto-thermopower (*MTEP*). For the *P* = 1600 W sample, the magnitude of *MR* reaches a peak value of 55 % around 245 K and shows a gradual decrease in the temperature range 200 to 50 K. At 50 K, MR is 40 %. In all other samples, the maximum value of MR is reached at 50 K. The sample for *P* = 1000 W exhibits the highest *MR* of 65 % at 50 K among all the samples. The *MTEP* is also maximum (~70 %) at 50 K among other samples. On the other hand, *MTEP* around the Curie temperature is highest (~85 %) for *P* = 1600 W and decreases to 40 % for *P* = 1200 W sample.

## 3.4. Magnetocaloric Effect

The magnetocaloric effect refers to a change in the temperature of a sample when it is magnetized/demagnetized and it is closely connected with a change in magnetic entropy during isothermal magnetization. Magnetic refrigeration based on magnetocaloric effect (MCE) make use of magnetocaloric materials is considered to be a promising alternative to the current conventional technology based on vapor compression in the near future because of its high cooling efficiency and non-use of global-warming gases [25]. Among the potential oxides for magnetic refrigeration, manganites are referred to as the most promising material because their tuneable ferromagnetic Curie temperature between 10 K and 375 K by adjusting compositions, high chemical stability, low production cost and reduced eddy current compared to metallic alloys because of their higher resistivity [26]. Magnetic entropy change $\Delta S_m = S_m(H)-S_m(H=0)$ is extracted from the magnetization isotherms (*M-H* isotherms) recorded at closed intervals.

We have studied the MCE for $LaMnO_3$ synthesised with MW powers $P = 1200$ W, 1400 W and 1600 W. Figure 4(a)-(c) show the $M(H)$ isotherms recorded at 5 K intervals over certain temperature range for each of the sample. Magnetic entropy change $\Delta S_m (H, T)$ is estimated from such isotherms using approximation to the Maxwell's equation $\Delta S_m (H,T) = \int_0^H \left(\frac{\partial M}{\partial T}\right)_H dH$ [26]. Figure 4(d-f) show the temperature dependence of $-\Delta S_m$ for $\Delta H = 10, 20$ and 30 kOe. The $-\Delta S_m$ exhibits a peak around $T_C$ for $P = 1600$ W and 1400 W samples and the magnitude of the peak increases with increasing $\Delta H$ value. On the other hand, $-\Delta S_m$ for $P = 1200$ W sample shows a plateau around $T \approx 125$ K - 200 K and decreases on either side. The maximum magnitude of $\Delta S_m$ for $\Delta H = 30$ kOe increases with increasing $P$ from ~0.83 J/kg.K for $P = 1200$ W to 2.21 J/kg.K for $P = 1400$ and to 4.78 J/kg.K for $P = 1600$ W. The obtained maximum $\Delta S_{max}$ value for $P = 1600$ W sample is higher than that of solid-state prepared $LaMnO_3$ (= 0.23 J kg$^{-1}$ K$^{-1}$ for $\Delta H = 30$ kOe [27]) and nanocrystalline (200-40 nm) $LaMnO_3$ ($\Delta S_m = 2.4$ -2.6 J kg$^{-1}$ K$^{-1}$ for $\Delta H = 50$ kOe) synthesised by sol-gel method [28] but comparable to nanocrystalline off-stoichiometric $LaMnO_3$ ($\Delta S_m \sim 4.8$-5.0 J kg$^{-1}$ K$^{-1}$ for average particle size of 100 nm) obtained for $\Delta H = 30$ kOe [29]. However, the obtained value in our sample is higher than other hole-doped perovskite manganites for $\Delta H = 30$ kOe [30,31,32,33]. While high values of $\Delta S_m$ and $\Delta T_{ad}$ are the primary quantities to establish a material as a good magnetic refrigerant, a large refrigerant capacity (RCP) is desirable for applications as it characterizes the temperature window for efficient operation of the magnetic Carnot cycle. The RCP is calculated using the relation, $RCP = |\Delta S_m^{max}| \times \delta T_{FWHM}$, where $\delta T_{FWHM}$ is the operating temperature range or full width at half maxima of the $\Delta S_m$ vs $T$ plot. The sample prepared at $P$ = 1400 W (125.5 J kg$^{-1}$) offers a higher RCP value than $P$ = 1600 W (102.5 J kg$^{-1}$). We could not estimate the RCP for $P$ = 1200 W as it shows a broad plateau in the $\Delta S_m$ vs $T$.

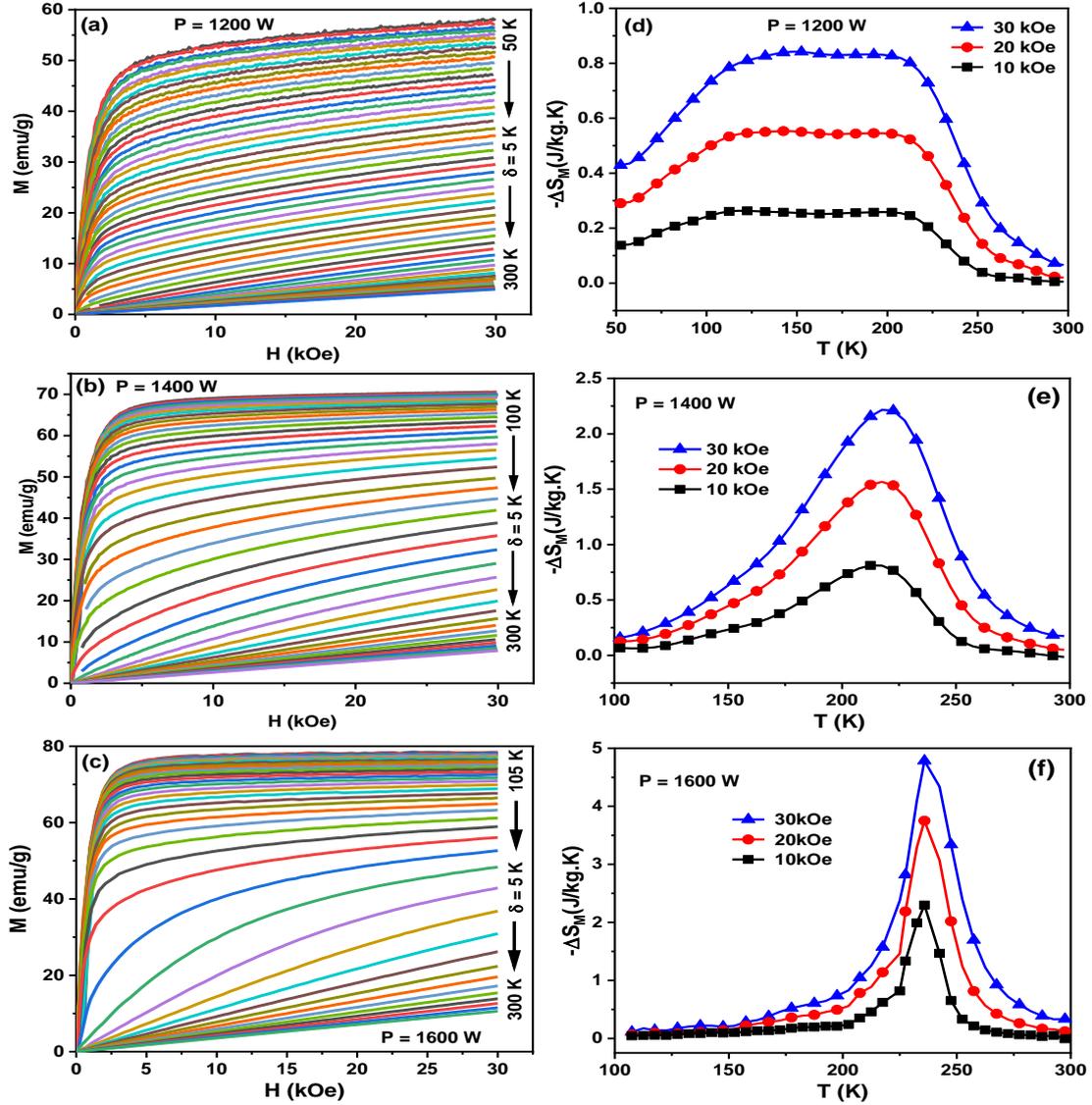

**Fig. 4.** Isothermal magnetization at 5 K intervals **(a-c)** and temperature dependence of the magnetic entropy change (*ΔS$_M$*) under different magnetic fields **(d-f)** for the LaMnO$_3$ prepared with MW $P$ = 1200 W, 1400 W and 1600 W, respectively.

## 3.5. Magnetostriction

Parallel magnetostriction ($\lambda_{par}$) represents the fractional change in the length experienced by a sample [$\lambda_{par} = [l(H)-l(H=0)]/l(H=0)$] upon magnetizing at fixed temperature and the strain is measured along the direction of the applied field. For randomly oriented grains, the overall change may be in volume ($\omega_{vol} = \lambda_{par} + 2\lambda_{per}$) or shear-type distortion ($\lambda_{anis} = \lambda_{par} - \lambda_{per}$) where $\lambda_{per}$ is the magnetostrain measured in a perpendicular direction to the applied magnetic field. It is found that in many Ca-based manganites, the magneto volume effect is dominant around $T_C$ and in the paramagnetic phase, and the magneto volume effect subdues below $T_C$. It was suggested that charges are localized as small lattice polarons with ferromagnetic clouds around them known collectively as magnetic polarons in zero external fields in the paramagnetic state, which leads to excess volume over the normal thermal expansion behavior [34,35]. Under an external magnetic field, magnetic polarons expand in size and charge becomes delocalized over a larger radius and the volume decreases, leading to a negative magnetovolume effect. In our dilatometer, we can measure the length change only along the direction of the applied field. The inset of Fig. 5(a) & (b) show the field dependence of magnetostriction at selected temperatures for $P$ = 1400 W and 1600 W samples. Referring to the inset of Fig. 5(a), we note that the magnetostriction is negative at 250 K and its magnitude smoothly increases with the applied magnetic field. At lower temperatures, a positive component develops at low fields which increases in amplitude and the contraction at higher fields decreases with decreasing temperature. We extract the maximum value at the highest field ($H$= 50 kOe) from the isothermal magnetostriction data and plot them in the main panels of Fig. 5(a) and (b) for $P$ = 1400 W and 1600 W samples, respectively. For $P$ = 1600 W sample, $\lambda_{par}$ is positive and small (-15 ppm) at 300 K and it increases in magnitude rapidly as the $T_C$ is approached. It reached a maximum magnitude (~180 ppm) at $T_C$ and rapidly decreases with temperature decreasing below $T_C$. The sign of $\lambda_{par}$ at the higher field crosses over to positive below 180 K, which is partially due to the rapid increase of the low-field behavior. The temperature dependence of $\lambda_{par}$ for $P$ = 1400 W looks similar to that of the $P$ = 1600 W sample. At 10 K, the magnetostriction for both $P$ = 1600 W and 14000 W are comparable (~ 40-50 ppm), however, the maximum magnitude of $\lambda_{par}$ around $T_C$ is 40 microstrain for $P$ = 1400 W which is at least four times smaller than $P$ = 1600 W sample. The larger value of magnetostriction in the $P$ = 1600 W sample is obviously related to the paramagnetic insulator-ferromagnetic metal transition, which is absent in the $P$ = 1400 W sample.

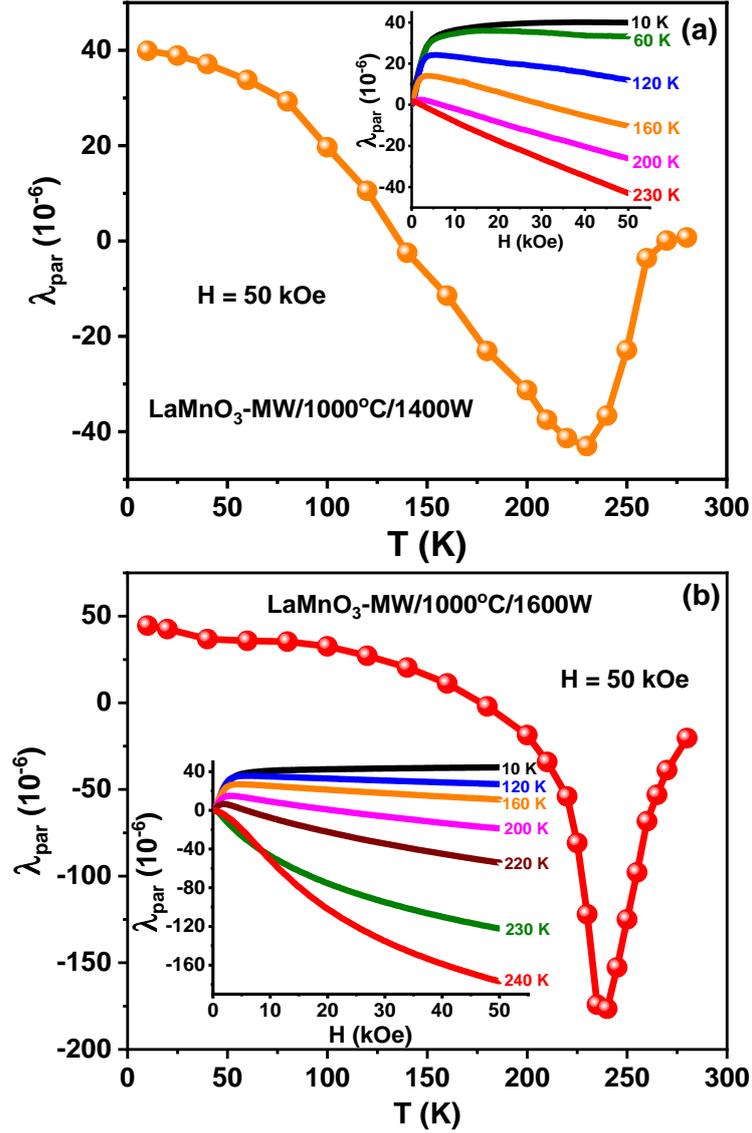

**Fig. 5.** Temperature dependence of $\lambda_{par}$ at $H$ = 50 kOe extracted from the field dependence of $\lambda_{par}$ data shown in the respective insets for LaMnO$_3$ prepared with $P$ = **(a)** 1400 W and **(b)** 1600 W.

## 4. Summary

We studied the influence of microwave power on magnetism, electrical resistivity, thermopower, magnetocaloric, magnetostrictive and spin resonance properties in LaMnO$_3$. With increasing MW power, Curie temperature increases, ferromagnetic transition becomes sharper, and an insulator-metal transition is induced with a sharp feature in the thermopower. Magnetoresistance, magnetic entropy change and magnetostriction values around $T_C$ were higher in the sample exposed to the maximum MW power ($P$ = 1600 W). Electron spin resonance study confirms that hole density increases with the MW power. Our study shows that tuning magnetism and other physical properties by the MW power is an easy and alternative route to achieve desirable properties without aliovalent or monovalent cation doping at the rare earth site in manganates.

## ACKNOWLEDGEMENTS

R. M. acknowledges the Ministry of Education, Singapore, for supporting this work (Grants numbers: A-0004212-00-00, A-8000462-00-00 and A-8000924-00-00)

## DATA AVAILABILITY

The data that support the findings of this study are available from the corresponding author upon reasonable request.